%% file: main.tex
\documentclass[manuscript,screen]{acmart}
\input{utils/packageref.tex}

\input{utils/solidity_listings_style}

\begin{document}

\title{LLM-Powered Detection of Price Manipulation in DeFi}

\author{Lu Liu}
\email{lliubf@connect.ust.hk}
\affiliation{%
  \institution{The Hong Kong University of Science and Technology}
  \department{Computer Science and Engineering Department}
  \city{Hong Kong}
  \country{China}
}

\author{Wuqi Zhang}
\email{wuqi.zhang@connect.ust.hk}
\affiliation{
    \institution{MegaETH}
    \city{Hong Kong}
    \country{China}
}

\author{Lili Wei}
\email{lili.wei@mcgill.ca}
\affiliation{
    \institution{McGill University}
    \city{Montreal}
    \country{Canada}
}

\author{Hao Guan}
\email{hguandl@icloud.com}
\affiliation{
    \institution{Nankai University}
    \city{Tian Jin}
    \country{China}
}

\author{Yongqiang Tian}
\email{yongqiang.tian@monash.edu}
\affiliation{
    \institution{Monash University}
    \city{Melbourne}
    \country{Australia}
}

\author{Yepang Liu}
\email{liuyp1@sustech.edu.cn}
\affiliation{
    \institution{Southern University of Science and Technology}
    \city{Shen Zhen}
    \country{China}
}

\author{Shing-Chi Cheung}
\email{scc@cse.ust.hk}
\affiliation{
    \institution{The Hong Kong University of Science and Technology}
    \city{Hong Kong}
    \country{China}
}

\input{sections/0_Abstract}

\maketitle

\input{sections/1_Introduction}
\input{sections/2_Background}
\input{sections/3_MotivationAttackModel}
\input{sections/4_Methodology}
\input{sections/5_Evaluation}
\input{sections/6_Discussion}

\input{sections/7_RelatedWork}
\input{sections/8_Conclusion}


\bibliographystyle{ACM-Reference-Format}
\bibliography{ref}

\end{document}

%% file: utils/packageref.tex
\usepackage{amsmath,amssymb,amsfonts}
\usepackage{graphicx}
\usepackage{mdframed}
\usepackage{textcomp}
\usepackage{xcolor}
\usepackage{mathtools}
\usepackage{listings}
\usepackage{xspace}
\usepackage{threeparttable}
\usepackage{enumitem}
\usepackage[most]{tcolorbox}
\tcbuselibrary{skins}
\def\BibTeX{{\rm B\kern-.05em{\sc i\kern-.025em b}\kern-.08em
    T\kern-.1667em\lower.7ex\hbox{E}\kern-.125emX}}

\usepackage[ruled,lined,linesnumbered,vlined]{algorithm2e}

\SetKwProg{Fn}{Function}{:}{}
\SetKwInOut{Input}{Input}
\SetKwInOut{Output}{Output}
\SetKw{Continue}{continue}
\SetKw{Break}{break}
\SetKwFunction{AlgAppend}{append}
\SetKwFunction{AlgAdd}{add}

\SetKwFunction{AlgTaintAnalysis}{TaintAnalysis}
\SetKwFunction{AlgPreProcess}{PreProcess}
\SetKwFunction{AlgIdentifySources}{IdentifySources}
\SetKwFunction{AlgPropagate}{Propagate}
\SetKwFunction{AlgIsSink}{IsSink}
\SetKwFunction{AlgReconstructPath}{ReconstructPath}

\SetKwData{AlgContract}{C}
\SetKwData{AlgCFGs}{CFGs}
\SetKwData{AlgTaintMap}{TaintMap}
\SetKwData{AlgOldTaintMap}{TaintMap$_{\text{old}}$}
\SetKwData{AlgInstruction}{Inst}
\SetKwData{AlgPath}{Path}

\SetKwArray{AlgTaintPaths}{TaintPaths}

\makeatletter
\renewcommand{\nllabel}[1]
 {{\let\@currentlabel\algocf@currentlabel
  \let\@currentcounter\algocf@currentcounter
  \label{#1}}}%

\renewcommand{\algocf@nl@sethref}[1]{%
  \renewcommand{\theHAlgoLine}{\thealgocfproc.#1}%
  \hyper@refstepcounter{AlgoLine}%
  \gdef\algocf@currentlabel{#1}%
  \gdef\algocf@currentcounter{AlgoLine}%
 }%
\makeatother

\newif\ifshowcomments
\showcommentsfalse 

\ifshowcomments
    \usepackage{fontawesome}
    \newcommand{\editnote}[3]{\xspace\colorbox{#1}{\sffamily \smaller \textcolor{white}{~\faCommenting{}~#2~}}\textcolor{#1}{~#3}\xspace}
    
    \newcommand{\todoc}[2]{{\textcolor{#1}{#2}}}
\else
    \newcommand{\editnote}[3]{}
    
    \newcommand{\todoc}[2]{}
\fi

\definecolor{lightgray}{rgb}{0.95, 0.95, 0.95}
\definecolor{lightgrey}{HTML}{F0F0F0}
\definecolor{darkgray}{rgb}{0.4, 0.4, 0.4}
\definecolor{editorGray}{rgb}{0.95, 0.95, 0.95}
\definecolor{editorOcher}{rgb}{1, 0.5, 0} 
\definecolor{editorGreen}{rgb}{0, 0.5, 0} 
\definecolor{orange}{rgb}{1,0.45,0.13}
\definecolor{olive}{rgb}{0.17,0.59,0.20}
\definecolor{brown}{rgb}{0.69,0.31,0.31}
\definecolor{purple}{rgb}{0.38,0.18,0.81}
\definecolor{lightblue}{rgb}{0.1,0.57,0.7}
\definecolor{lightred}{rgb}{1,0.4,0.5}
\definecolor{codegreen}{rgb}{0,0.6,0}
\definecolor{codegray}{rgb}{0.5,0.5,0.5}
\definecolor{codepurple}{rgb}{0.58,0,0.82}
\definecolor{backcolour}{rgb}{0.95,0.95,0.92}
\definecolor{maroon}{rgb}{0.5, 0.0, 0.0}

\definecolor{commentbox}{HTML}{C68A75}
\definecolor{responsebox}{HTML}{A3BE8C}

\newcommand{\todoblue}[1]{\todoc{blue}  {[#1]}}

\newcommand{\todocyan}[1]{\todoc{cyan}  {[#1]}}

\newcommand{\todopurple}[1]{\todoc{purple}  {[#1]}}

\newcommand{\lu}[1]{\todopurple{Liulu: #1}}

\newcommand{\lili}[1]{\todoblue{Lili: #1}}
\newcommand{\yepang}[1]{\todocyan{Yepang: #1}}

\definecolor{blue1}{HTML}{385391}
\definecolor{blue2}{HTML}{566DA2}
\definecolor{blue3}{HTML}{7F92BA}
\definecolor{blue4}{HTML}{B2BED8}
\definecolor{blue5}{HTML}{D5DCED}
\definecolor{blue6}{HTML}{DBE1F0}

\definecolor{promptLowerBg}{HTML}{DFE9F6}
\definecolor{promptupperBg}{HTML}{F2F2F2}

\definecolor{boxbg}{RGB}{240, 240 ,240}

\newenvironment{answertorq}{%
    \begin{tcolorbox}[
            arc=2mm,
            boxrule=0.5pt,
            left=2pt,
            right=2pt,
            top=2pt,
            bottom=2pt
        ]
}{\end{tcolorbox}}

\definecolor{codegreen}{rgb}{0,0.6,0}
\definecolor{codegray}{rgb}{0.5,0.5,0.5}
\definecolor{codeblue}{rgb}{0,0,1}
\definecolor{codeteal}{rgb}{0.16, 0.5, 0.44}
\definecolor{codepurple}{rgb}{0.5,0.0,0.5}
\definecolor{codebackground}{rgb}{0.96,0.96,0.96}

\lstdefinestyle{soliditystyle}{
    backgroundcolor=\color{codebackground},
    commentstyle=\color{codegreen},
    keywordstyle=[1]{\bfseries\color{codeblue}},
    keywordstyle=[2]{\bfseries\color{codeteal}},
    keywordstyle=[3]{\bfseries\color{codepurple}},
    numberstyle=\small\color{black},
    basicstyle=\ttfamily\small,
    breaklines=true,
    breakatwhitespace=false,
    captionpos=b,
    keepspaces=true,
    numbers=left,
    numbersep=5pt,
    showspaces=false,
    showstringspaces=false,
    showtabs=false,
    tabsize=2,
    comment=[l]{//},
    morekeywords={
        [1]function, public, private, internal, external, returns, return, interface, constructor, modifier, view, pure, payable, constant, if, else, for, while, break, require, assert, revert, event, emit, new, delete, using, is, try, catch 
        [2]address, bytes, memory, string, storage, calldata, bool, int, unit, unit256, mapping, struct, enum    
        [3]delegatecall, call, staticcall, msg, tx, block, this, super, selfdestruct, balance        
    }
}

\newmdenv[
  backgroundcolor=lightgrey,
  linecolor=lightgrey,
  innerleftmargin=10pt,
  innerrightmargin=10pt,
  innertopmargin=10pt,
  innerbottommargin=10pt,
  skipabove=10pt,
  skipbelow=10pt,
  linewidth=0pt,
  frametitleaboveskip=10pt,
  frametitlealignment=\centering,
  frametitlerule=false
]{llmprompt}

\newcommand{\eg}{\mbox{e.g.}\xspace}

\newcommand{\ie}{\mbox{i.e.}\xspace}


%% file: utils/solidity_listings_style.tex
\usepackage{xcolor}

\definecolor{codegreen}{rgb}{0,0.6,0}
\definecolor{codegray}{rgb}{0.5,0.5,0.5}
\definecolor{codepurple}{rgb}{0.58,0,0.82}
\definecolor{backcolour}{rgb}{0.97,0.97,0.94}

\definecolor{keywordblue}{rgb}{0.13,0.13,0.81}     
\definecolor{typeorange}{rgb}{0.8,0.33,0.0}        
\definecolor{globalviolet}{rgb}{0.5,0.0,0.5}       
\definecolor{userteal}{rgb}{0.0,0.5,0.5}           

\lstdefinelanguage{Solidity}{
    morekeywords={public, private, internal, external, view, pure, 
                  payable, constructor, event, emit, require, revert}, 
    morekeywords=[2]{address, bool, bytes, string, uint, uint8, uint16, uint32, uint64, uint128, uint256, int, int8, int16, int32, int64, int128, int256, memory, storage, calldata}, 
    morekeywords=[3]{block, block.chainid, block.coinbase, block.difficulty, block.gaslimit, block.number, block.timestamp,  msg.data, msg.sender, msg.sig, msg.value, tx, tx.gasprice, tx.origin, this}, 
    morekeywords=[4]{SafeMath, ERC20, IERC20},   
    sensitive=true,
    comment=[l]{//},
    morecomment=[s]{/*}{*/},
    stringstyle=\color{red!80!black},
}

\lstdefinestyle{soliditystyle}{
    backgroundcolor=\color{gray!10},
    commentstyle=\color{codegreen},
    keywordstyle=\color{keywordblue},              
    keywordstyle=[2]\color{typeorange},            
    keywordstyle=[3]\color{globalviolet},          
    keywordstyle=[4]\color{userteal},             
    numberstyle=\tiny\color{codegray},
    stringstyle=\color{red!80!black},
    basicstyle=\ttfamily\footnotesize\color{black}, 
    breakatwhitespace=false,
    breaklines=true,
    captionpos=b,
    keepspaces=true,
    numbers=left,
    numbersep=5pt,
    showspaces=false,
    showstringspaces=false,
    showtabs=false,
    tabsize=2,
}

\usepackage{cleveref} 

\Crefname{algocf}{Algorithm}{Algorithms}
\crefname{algocf}{Algorithm}{Algorithms}

\Crefname{algorithm}{Algorithm}{Algorithms}
\crefname{algorithm}{Algorithm}{Algorithms}

\crefname{appendix}{Appendix}{Appendices}
\Crefname{appendix}{Appendix}{Appendices}

\Crefname{figure}{Figure}{Figures}
\crefname{figure}{Figure}{Figures}

\crefname{listing}{Listing}{Listings}
\Crefname{listing}{Listing}{Listings}

\Crefname{table}{Table}{Tables}
\crefname{table}{Table}{Tables}

\crefname{thm}{Theorem}{Theorems}
\Crefname{thm}{Theorem}{Theorems}

\crefname{equation}{Equation}{Equations}
\Crefname{equation}{Equation}{Equations}

\crefformat{chapter}{\S~#2#1#3}
\crefmultiformat{chapter}{\S\S~#2#1#3}{ and~#2#1#3}{, #2#1#3}{, and~#2#1#3}

\crefformat{section}{\S~#2#1#3}
\crefmultiformat{section}{\S\S~#2#1#3}{ and~#2#1#3}{, #2#1#3}{, and~#2#1#3}

%% file: sections/0_Abstract.tex
\begin{abstract}
Decentralized Finance (DeFi) smart contracts manage billions of dollars, making them a prime target for exploits. Price manipulation vulnerabilities, often via flash loans, are a devastating class of attacks causing significant financial losses. Existing detection methods are limited. Reactive approaches analyze attacks only after they occur, while proactive static analysis tools rely on rigid, predefined heuristics, limiting adaptability. Both depend on known attack patterns, failing to identify novel variants or comprehend complex economic logic. We propose PMDetector, a hybrid framework combining static analysis with Large Language Model (LLM)-based reasoning to proactively detect price manipulation vulnerabilities. Our approach uses a formal attack model and a three-stage pipeline. First, static taint analysis identifies potentially vulnerable code paths. Second, a two-stage LLM process filters paths by analyzing defenses and then simulates attacks to evaluate exploitability. Finally, a static analysis checker validates LLM results, retaining only high-risk paths and generating comprehensive vulnerability reports. To evaluate its effectiveness, we built a dataset of 73 real-world vulnerable and 288 benign DeFi protocols. Results show PMDetector achieves 88\% precision and 90\% recall with Gemini 2.5-flash, significantly outperforming state-of-the-art static analysis and LLM-based approaches. Auditing a vulnerability with PMDetector costs just \$0.03 and takes 4.0 seconds with GPT-4.1, offering an efficient and cost-effective alternative to manual audits.
\end{abstract}

%% file: sections/1_Introduction.tex
\section{Introduction}
Decentralized Finance (DeFi)~\cite{defi} has emerged as a significant force on the blockchain, with automated smart contracts managing over a hundred billion dollars in value~\cite{defiaum}.
These contracts enable novel financial services such as decentralized exchanges (DEX) and lending protocols~\cite{jensen2021introduction}.
However, core blockchain principles such as immutability and autonomous execution make it crucial to ensure the security of smart contracts. Once deployed, a flawed contract cannot be easily patched.
A single vulnerability can be exploited systematically, leading to catastrophic financial losses~\cite{top100defihack,top10attack}.

Among diverse security threats, price manipulation vulnerabilities are particularly critical and financially devastating.
These attacks occur when an adversary artificially distorts an asset's price as reported by on-chain price oracles~\cite{zhang2023demystifying}. 
Victim protocols that blindly trust the price information for critical operations, such as determining collateral value or calculating swap rates, become vulnerable to exploitation.
Attacks recorded on DefiHacklabs~\cite{defihacklabs} between Jan 2023 and May 2025 show that price manipulation was the root cause of 17.3\% of all major exploits, resulting in losses exceeding \$165.8 million~\cite{defiincidentlist}.
BonqDao, one of the highest-impact price manipulation attacks, caused a loss of \$88.0 million~\cite{bonqdao}.
A common attack vector involves an attacker using a flash loan to execute a large trade, which temporarily unbalances liquidity pools on decentralized exchanges. A victim protocol using the spot price from pools as an oracle can then be drained. 

The research community has developed automated detection techniques, broadly categorized into post-mortem and pre-mortem analysis approaches~\cite{wu2023defiranger, xie2024defort, kong2023defitainter, wu2024strengthening, bosi2025following}.
Post-mortem analysis tools like DeFiRanger~\cite{wu2023defiranger} and DeFort~\cite{xie2024defort} analyze on-chain transactions for already occurring attacks. While effective for monitoring attacks, they are not applicable for pre-deployment vulnerability prevention.
Conversely, static analysis frameworks like SMARTCAT~\cite{bosi2025following} and DeFiTainter~\cite{kong2023defitainter} offer a pre-mortem solution by proactively analyzing source code or bytecode.
However, their efficacy is constrained by reliance on predefined, manually-crafted heuristics. This rigidity restricts their ability to reason about semantic nuances.
They struggle to identify novel price manipulation vulnerabilities that deviate from known patterns, leading to critical false negatives (FNs).
This rigidity can also produce false positives (FPs) when encountering legitimate but unconventional defense mechanisms.
Thus, a critical gap exists: the need for proactive detection that transcends fixed patterns by combining systematic code analysis with deeper, contextual understanding of economic exploit logic.

The need for proactive and context-aware detection motivates exploring new paradigms beyond traditional program analysis.
The recent success of Large Language Models (LLMs) in code understanding and bug detection~\cite{nam2024using,li2024enhancing, ma2023lms,jin2024llms} presents a new opportunity for smart contract security analysis.
Recent studies~\cite{david2023needmanual, gao2024unveiling, wei2024llm} have demonstrated that models like GPT-4 could identify logic flaws in smart contracts. However, these approaches often suffer from high false positive rates, limiting practical applicability.
Subsequent research has focused on improving LLM-based detection reliability through combining LLM analysis with static analysis~\cite{hu2023large, sun2024gptscan, wu2024advscanner}, implementing multi-agent pipelines~\cite{wei2024llm, wei2025advanced}, and developing fine-tuned domain-specific LLMs~\cite{zhong2025defiscope, ma2024combining, bu2025enhancing}.
Despite demonstrated success on common vulnerabilities, these general-purpose LLM approaches often prove ineffective at identifying sophisticated economic exploits like price manipulation attacks. For example, GPTScan~\cite{sun2024gptscan} uses LLMs to match vulnerability patterns within individual functions; this function-centric view is inadequate for detecting price manipulation, which is a path-dependent exploit spanning multiple contract interactions.  

We summarize three main challenges for applying LLMs to the price manipulation detection task:
\textbf{C1: Lack of domain-specific attack knowledge.}
Price manipulation is fundamentally an economic exploit, not a conventional code bug. Detection requires a deep understanding of DeFi primitives such as AMM invariants and oracle mechanics. General-purpose LLMs, trained primarily on vast code corpora~\cite{jiang2024survey}, lack this specialized financial knowledge and cannot distinguish legitimate trades from manipulative ones.
\textbf{C2: Inherent hallucination in code reasoning.}
LLMs are probabilistic systems rather than logical execution engines~\cite{vaswani2017attention}. They generate explanations by predicting likely word sequences rather than simulating code execution.
This makes them frequently ``hallucinate'' execution paths that are syntactically plausible but semantically infeasible~\cite{sriramanan2024llm}, undermining reliability for security analysis.
\textbf{C3: Limited generalization beyond seen patterns.}
Current models exhibit over-reliance on syntactic patterns from training data. They often fail to identify novel attack vectors that are semantically equivalent to known exploits but syntactically distinct.
Unlike well-documented vulnerabilities such as reentrancy~\cite{commonattack}, price manipulation vulnerabilities suffer from limited representation in existing datasets, further constraining LLM understanding.
\yepang{We should also mention LLM-based smart contract vulnerability detection approaches here and argue that they are not capable of effectively detecting price manipulations.} \lu{Updated.}

To address these challenges, we present \texttt{PMDetector}, a novel framework that integrates taint analysis, LLM-based reasoning, and rule-based static checking. 
First, to address the lack of domain knowledge (C1) and poor generalization (C3), we introduce a formal attack model that defines price manipulation semantics. 
The model guides static taint analysis to identify data flows where external price input could influence critical financial operations, and guides LLM prompt design for path filtering and attack simulation.
By constraining the problem space, the model focuses LLM analysis on plausible vulnerabilities and relevant DeFi exploit context, reducing false negatives.
Next, we leverage a two-stage LLM pipeline as a semantic analysis engine. 
In the path filtering stage, the LLM evaluates defense mechanism efficacy, pruning safe execution flows. For remaining high-risk candidates, the attack simulation prompts the LLM to construct viable exploit scenarios. This structured pipeline concentrates LLM reasoning capabilities and lowers the risk of false positives.
Finally, to mitigate LLM hallucination (C2), a rule-based static checker validates remaining high-risk paths. This module acts as a lightweight verifier, cross-referencing paths against established defense mechanisms, eliminating false positives from LLM hallucinations.
\yepang{How do we perform checks? Using rules?}\lu{Yes, we use pre-defined rules. I've rephrased this part.}

To evaluate \texttt{PMDetector}, we construct a benchmark of real-world DeFi protocols with confirmed price manipulation vulnerabilities. The benchmark comprises 73 contracts from prior benchmarks~\cite{xie2024defort, chen2024flashsyn, wu2023defiranger, wu2024strengthening} and the public security dataset DeFiHackLabs~\cite{defiincidentlist}.
The set spans diverse protocol categories and covers incidents from 2020 to 2025, providing broad coverage of price-manipulation patterns. Financial losses from these 73 contracts exceed \$258 million~\cite{defihacklabs}. 
\yepang{Are the contracts diversified? Are the attacks caused by different kinds of defects in code? Say something impressive about the dataset.}\lu{Updated.}
We also include a balanced set of 288 benign contracts from prior studies~\cite{kong2023defitainter}.
Results demonstrate that \texttt{PMDetector} achieves 88\% precision and 90\% recall using Gemini 2.5-flash~\cite{gemini2.5flash}, outperforming state-of-the-art static analysis and LLM-based baselines. 
Ablation studies confirm the contribution of each component and removing any component leads to substantial drops in precision and recall, validating our architectural design. 
The cost to audit a price manipulation vulnerability with \texttt{PMDetector} is \$0.03 for GPT 4.1, substantially lower than typical human-audit costs (\$5,000-\$15,000)~\cite{auditcost}.

In summary, this paper makes the following contributions.
\begin{itemize}[topsep=0pt, leftmargin=*]
    \item \textbf{A comprehensive benchmark for price manipulation.}
    We construct a comprehensive benchmark comprising 73 real-world vulnerable contracts responsible for over \$258 million in losses. Systematically curated from prior research and public incident reports, the benchmark offers broad coverage of attack patterns across diverse DeFi protocols from 2020 to 2025.
    \item \textbf{A novel hybrid detection framework.}
    We present \texttt{PMDetector}, a hybrid framework combining static analysis and LLM-based reasoning to overcome individual limitations. It employs static taint analysis for candidate identification and performs a two-step LLM pipeline for context-aware semantic validation, followed by static validation checking.
    \item \textbf{Superior detection performance.}
    We evaluate \texttt{PMDetector} on our benchmark and safe real-world DeFi protocols. Results demonstrate state-of-the-art accuracy, significantly outperforming existing static analysis and LLM-based baselines.
\end{itemize}


%% file: sections/2_Background.tex
\section{Background}
\subsection{Automated Market Makers and Price Discovery in DeFi Ecosystem}
Automated Market Makers (AMMs)~\cite{cpmm} emerged as a solution to address the fundamental limitations of traditional Central Limit Order Books (CLOBs)~\cite{clob} in on-chain transaction environments.
Traditional CLOBs operate through a centralized matching mechanism where market participants submit buy and sell orders at specified price levels, which are subsequently matched by a central authority.
This approach faces significant scalability challenges when deployed in decentralized environments, particularly due to high transaction costs and substantial network latency.
AMMs restructure decentralized trading by replacing the order book mechanism with liquidity pools, which are smart contracts holding reserves of two or more tokens. Liquidity providers (LPs)~\cite{lp} deposit paired assets into these pools and receive LP tokens proportional to their contribution to the pool's total liquidity. These pooled assets are then available for other users to trade against.

The core of an AMM lies in its pricing algorithm, which determines the exchange rate between the assets in the pool based solely on their respective quantities. The most influential and widely adopted model is the constant product market maker (CPMM)~\cite{cpmm}.
A CPMM maintains the invariant $x \cdot y = k$, where $x$ and $y$ represent the quantities of two tokens (e.g., ETH~\cite{eth} and USDC~\cite{udsc}) in the pool, and $k$ is a constant.
When users initiate a trade, they deposit a certain amount of one token ($\Delta x$) into the pool and receive a corresponding amount of the other token ($\Delta y$), ensuring that the invariant $(x+\Delta x) \cdot (y - \Delta y) = k$ remains satisfied (excluding transaction fees).

The spot price in such a pool is not explicitly stored but rather derived implicitly as the ratio of the reserves: $P = x/y$. Consequently, any transaction that alters the reserve quantities $x$ and $y$ simultaneously modifies the instantaneous price.
This mechanism enables on-chain price discovery and serves as a critical primitive in the decentralized finance (DeFi) ecosystem~\cite{defi}.
The price information reported by prominent AMM pools is often used by other decentralized applications (DApps)~\cite{dapp}, such as lending protocols, derivatives platforms, and yield aggregators. These DApps rely on AMM-derived prices as price oracles to determine collateral values, calculate liquidation thresholds, and settle financial contracts.
However, the reliance on AMM-derived pricing creates a significant attack vector that can be exploited by malicious actors.


\subsection{Price Manipulation Vulnerabilities in DeFi Protocols}
Since AMM prices are direct functions of token reserves within liquidity pools, actors with sufficient capital can execute large trades that significantly alter reserve ratios, creating temporary but substantial price dislocation.
The goal of such an attack is not to profit from the trade itself, but to exploit a separate, victim protocol that uses the manipulated price from the AMM as a trusted oracle.
Historically, executing such an attack required an immense upfront capital commitment, making it impractical for all but the wealthiest actors. However, the introduction of flash loans~\cite{defi} fundamentally transformed this attack landscape.
A flash loan is an uncollateralized loan that must be borrowed and repaid within the same atomic transaction~\cite{flashloan}.
If the borrower cannot repay the full amount plus a small fee by the end of the transaction, the entire transaction, including all actions performed, is reverted.
This mechanism allows an attacker to borrow millions of dollars' worth of cryptocurrency for a few seconds, use it to manipulate market conditions, and repay the loan using the proceeds of their exploit, all with zero initial capital.

In this work, we focus on flash loan-based vulnerabilities, excluding front-running and sandwich attacks, narrowing the scope to manipulations occurring within one transaction, leveraging the atomic nature of blockchain transactions.



%% file: sections/3_MotivationAttackModel.tex
\section{Motivation and Attack Model}
This section presents an example of real-world price manipulation vulnerabilities, followed by the definition of an attack model that captures the essential characteristics and attack vectors of this vulnerability class.

\subsection{Illustrating Example}
\begin{figure}[t!]
\small
\begin{lstlisting}[xleftmargin=2em,]
function burnToHolder(uint256 amount, address sender) external {
    address[] memory path = new address[](2);
    path[0] = address(_burnToken);  // ZongZi token
    path[1] = uniswapRouter.WETH();  // WBNB
    uint256 deserved = 0;
    deserved = uniswapRouter.getAmountsOut(amount, path)[path.length - 1];
    require(payable(address(_burnToken)).balance >= deserved);
    _burnToken.zongziToholder(sender, amount, deserved);
    burnFeeRewards(sender, deserved);
}

function burnFeeRewards(address to, uint256 increase) private {
    address sender = to;
    _transfer(address(this), sender, increase);
    burnAmount[sender] = burnAmount[sender].add(increase);
}
\end{lstlisting}
\caption{Vulnerable logic in the \texttt{ZZF} protocol.}
\label{lst:zzf_burn}
\end{figure}

\cref{lst:zzf_burn} shows a real-world price manipulation vulnerability found in the \texttt{ZZF} protocol~\cite{zzfcontract}, a BEP-20 token contract deployed on the BNB Smart Chain~\cite{bsc}. The exploit resulted in a loss of approximately \$223,000~\cite{zzfloss}.
The vulnerability resides within the \texttt{ZZF} smart contract's reward distribution mechanism. The protocol rewards users with \texttt{ZZF} tokens for burning \texttt{ZongZi} tokens (\texttt{\_burnToken} at line 3), with rewards proportional to the burned tokens' value. The flaw lies in how the contract determines this value.
The \texttt{burnToHolder} function uses the spot price from a PancakeSwap liquidity pool~\cite{pancakeswap} as its price oracle via \texttt{uniswapRouter.getAmountsOut} (line 6). The  variable \texttt{deserved}, representing the burned tokens' value in WBNB, is calculated based on current reserves of the ZongZi/WBNB liquidity pool.
In automated market makers (AMMs), token spot prices are determined by reserve ratios within liquidity pools. The spot price of WBNB is $x/y$, where $x$ and $y$ represent WBNB and ZongZi token reserves, 
subject to the constant product constraint $x \cdot y = k$. This mechanism makes both prices highly susceptible to manipulation through large trades that alter the reserve ratio.
The manipulated \texttt{deserved} value is passed to \texttt{burnFeeRewards}, which calculates and distributes the \texttt{ZZF} reward (line 9). This function uses the manipulated \texttt{increase} value to transfer disproportionately large amounts of \texttt{ZZF} tokens to the attacker (line 14). The attacker then calls \texttt{receiveRewards} to swap these tokens for WBNB, draining the contract's funds.
\lili{I think the previous procedure is hard to understand possible because it is too high level and abstract.}\lu{I have updated it. See if it is better.}

Static analysis excels at detecting data flows but does not interpret business logic or economic intent. Consider the ZZF contract: a static analyzer can trace that \texttt{burnToHolder} invokes \texttt{burnFeeRewards}, which calls \texttt{\_transfer}, showing data flow from \texttt{getAmountsOut} to token transfer. However, this appears as disconnected operations, not a coherent economic mechanism. In contrast, an LLM can infer intent from function names like \texttt{burnToHolder} and \texttt{burnFeeRewards}, suggesting a reward mechanism, and variable names like \texttt{deserved}, implying fair token valuation. The LLM recovers the core business logic: rewarding users with ZZF tokens proportional to the burned ZongZi token value.
Once this intent is clear, the security implication follows: manipulating the price oracle undermines the contract's economic premise, enabling attackers to claim inflated, unearned rewards.

\subsection{Attack Model} \label{sec:attack-model}
\begin{figure*}[t!]
    \centering
    \includegraphics[width=5in]{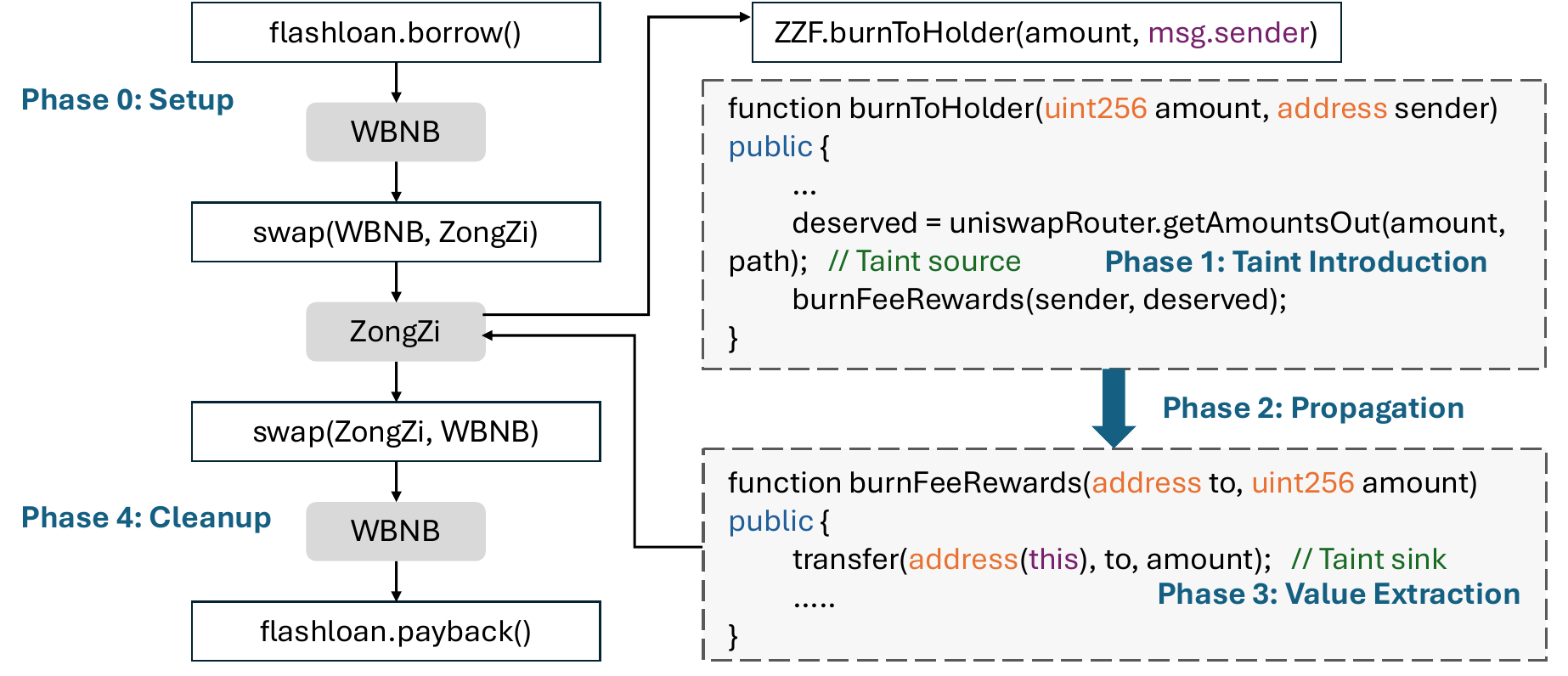}
    \caption{Demonstration of the Attack on the ZZF Protocol}
\label{lst:zzf_attack_model}
\end{figure*}

Following existing practices~\cite{wang2024defiguard, zhong2025defiscope, xie2024defort, wen2024foray}, our attack model formalizes flash-loan-based price manipulation as a multi-stage process, which begins with a flash loan and proceeds through manipulation, exploitation, value extraction, and cleanup.
We develop taint analysis and LLM reasoning prompts based on this model.
We use the ZZF protocol (\cref{lst:zzf_burn}) and its attack demonstration (\cref{lst:zzf_attack_model}) for illustration.

\textit{1) Phase 0: Setup.}
Focusing on flash-loan-based price manipulation, the attacker first acquires exploitable capital via a flash loan.
For example, in \cref{lst:zzf_attack_model}, an attacker initiates a flash loan to borrow large amounts of WBNB without collateral, serving as capital for subsequent manipulation.

\textit{2) Phase 1: Taint Introduction.}
The attacker introduces a tainted price into the DeFi ecosystem as the taint source, performing actions that cause the protocol's price-reporting mechanism to reflect abnormal values through large DEX swaps or direct oracle manipulation.
In the ZZF protocol, the attacker invokes \texttt{swap} to exchange WBNB for ZongZi tokens within the ZongZi/WBNB pool. This dramatically reduces WBNB reserves while inflating ZongZi reserves, artificially elevating ZongZi price relative to WBNB, resulting in abnormal ZongZi pricing when \texttt{uniswapRouter.getAmountsOut} is called.

\textit{3) Phase 2: Propagation and Exploitation.}
The attacker propagates the manipulated oracle price to subvert protocol logic. In \cref{lst:zzf_attack_model}, the attacker invokes \texttt{burnToHolder} within the ZZF protocol. This function queries the ZongZi/WBNB pool for ZongZi token price, receiving the inflated price from Phase 1 manipulation.
This manipulated price passes to \texttt{burnFeeRewards}, subverting core contract logic by using the malicious value to calculate rewards, transferring illegitimately large quantities of ZZF tokens to the attacker and corrupting internal accounting.
\lili{Is ``tainted data'' a good name? We taint this data to track their flow but they are tainted due to some other reasons (e.g., they are prices that can be manipulated). I think in this motivation, we may not call them tainted data}\lu{Agree. I have rephrased this part.}

\textit{4) Phase 3: Value Extraction.}
The attacker capitalizes on the inflated token balance to drain protocol funds through direct mechanisms (transferring borrowed assets) or indirect methods (updating the attacker's balance in internal accounting for subsequent withdrawal).
In ZZF protocol, the attacker calls \texttt{burnFeeRewards}, transferring large balances of unfairly minted ZZF tokens.

\textit{5) Phase 4: Cleanup.}
The final step cleans up positions and repays flash loans. The attacker unwinds positions, repays loans, restores oracles if necessary, and finalizes profit.
In \cref{lst:zzf_attack_model}, the attacker swaps remaining ZongZi tokens back to WBNB, restoring the liquidity pool's price.

Attackers may loop through these phases, \eg, by re-manipulating oracles after initial value extraction (circular attacks) or chaining exploits across protocols.
Our model effectively captures these complex scenarios by focusing on end-to-end flow from manipulated source to profitable sink, regardless of intermediate complexity.


%% file: sections/4_Methodology.tex
\section{Methodology}
\begin{figure*}[t!]
    \centering
    \includegraphics[width=5.5in]{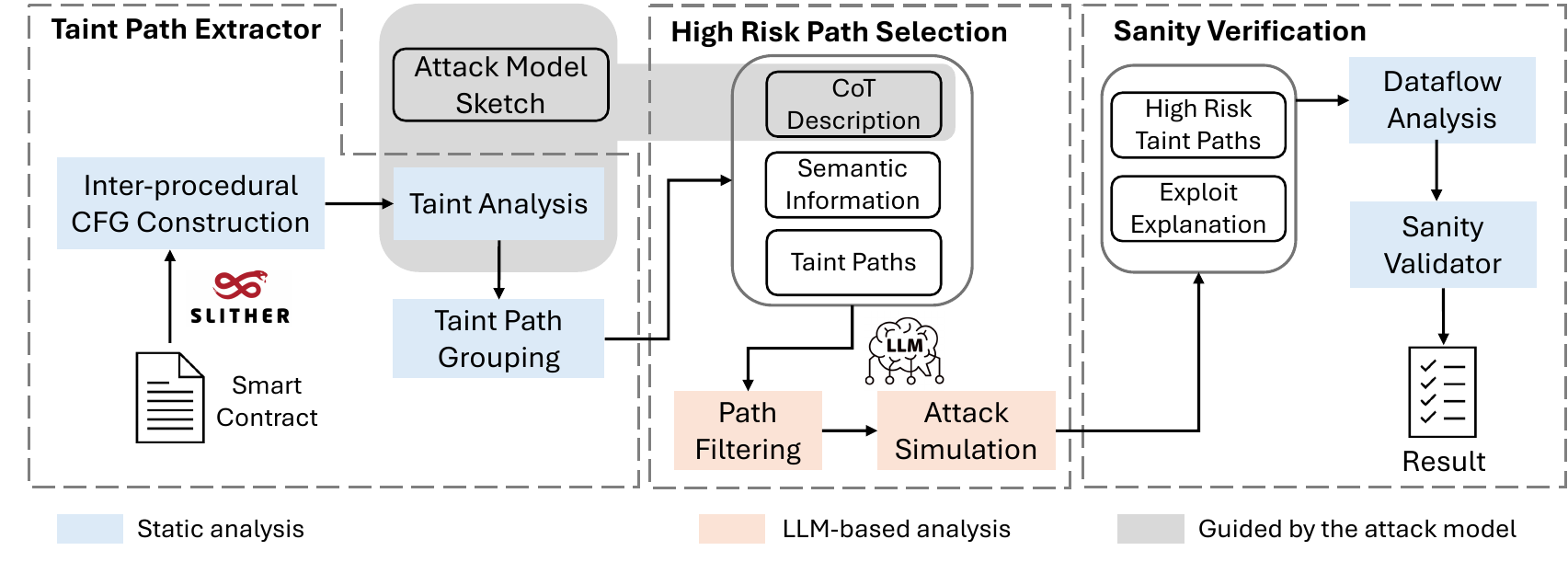}
    \caption{The workflow of \texttt{PMDetector}.
    }
    \label{method_workflow}
\end{figure*}

We present \texttt{PMDetector}, a hybrid framework integrating inter-procedural taint analysis, LLM-based reasoning, and semantic sanity checking to detect price manipulation vulnerabilities in DeFi smart contracts.
Our approach operates in three phases: (1) formal attack model-guided static taint analysis with path grouping to reduce token consumption, (2) a two-step LLM pipeline for defense validation and high-risk path selection, and (3) a rule-based semantic checker for false positive reduction.
Figure~\ref{method_workflow} illustrates \texttt{PMDetector}'s framework.

\subsection{Static Taint Analysis}
Based on the attack model in Section~\ref{sec:attack-model}, we propose an inter-procedural, flow-sensitive static taint analysis framework for identifying vulnerable paths in smart contracts using Slither Intermediate Representation (SlithIR)~\cite{feist2019slither}.
The target DeFi protocol is converted into an inter-procedural control-flow graph (ICFG), then we initiate taint analysis from identified sources and track taints by traversing the ICFG.
To overcome limitations of existing heuristic-based approaches~\cite{kong2023defitainter, wu2024strengthening, bosi2025following}, rather than relying on pattern matching, we analyze semantic and economic intent in code patterns for more accurate vulnerability detection.
We define semantic primitives extracted from the smart contract's IR (Table~\ref{tab:semantic_primitives}) as foundational facts for higher-level semantic analysis and establish inference rules based on these primitives.
The notation $v\downarrow$ represents that variable $v$ is tainted. A $Sink(op)$ predicate is derived if a tainted value reaches operation $op$.

\begin{table*}[t!]
  \caption{Definition of Semantic Primitives.}
  \label{tab:semantic_primitives}
  \small
  \begin{tabular}{p{0.18\linewidth} p{0.78\linewidth}}
    \toprule
    \textbf{Predicate} & \textbf{Description} \\
    \midrule
    \textit{Call(cs, f, args, ret)} &
    At call site \textit{cs}, function \textit{f} is called with arguments \textit{args}, and its result is assigned to \textit{ret}. \\
    \hline
    \textit{EC(cs, f, ...)} &
    A call to a function \textit{f} in an external contract. \\
    \hline
    \textit{SSTORE($\sigma$, v)} &
    The value of variable \textit{v} is written to storage slot \textit{$\sigma$} (\texttt{SSTORE}).\\
    \hline
    \textit{IsPublic(f)} &
    Function \textit{f} has \texttt{public} or \texttt{external} visibility.\\
    \hline
    \textit{Transfer(r, a)} &
    \textit{r} and \textit{a} are the recipient and the transfer amount for a transfer function. \\
    \hline
    \textit{Arg(cs, i, v)} &
    At call site \textit{cs}, the variable \textit{v} is passed as the i-th argument. \\
    \hline
    \textit{IsMappingSlot($\sigma$)} &
    A mapping from address \textit{addr} to a uint type \textit{unit} at base slot \textit{$\sigma$}. \\
    \bottomrule
    \end{tabular}
\end{table*}

\subsubsection{Taint Sources}
Taint sources are entry points for attacker manipulation, aligning with phase 1 of the attack model. We consider any external information potentially controlled by the attacker as a taint source.
Specifically, there are two kinds of taint sources, i.e., data directly provided by the attacker and data indirectly loaded from external contracts that may be controlled by the attacker.
The former kind includes inputs of \texttt{public} or \texttt{external} functions and transaction property variables  (e.g.~\texttt{msg.data}, \texttt{msg.value}).
The latter are typically external price oracle contracts. We conservatively identify such contracts by marking external function calls as taint sources if they are \texttt{view} or \texttt{pure} and their return values are used in multiplication or division operations. This pattern indicates price/ratio calculations and discovers custom-wrapped oracle calls.
We also mark return values from known DEX functions (e.g., \texttt{getAmountsOut}, \texttt{getReserves}) as taint sources. For example, in \cref{lst:zzf_burn}, \texttt{deserved} is identified as a taint source tainted by \texttt{uniswapRouter.getAmountsOut()}.

\subsubsection{Taint Propagation Rules}
Taint propagation rules define how taint spreads through data and control dependencies, modeling Phase 2 of our attack model. We consider four types of rules:

\textit{1) Intra-procedural Data Flow:}
Taint propagates through direct data dependencies within a function. If a variable's value is computed using one or more tainted operands (e.g., in an assignment or binary operation), this variable also becomes tainted.

\textit{2) Inter-procedural Data Flow:}
Taint flows across function call boundaries. When a tainted value is passed as an argument to a function, the corresponding parameter in the callee is tainted. Conversely, a tainted value returned from a function taints the variable that receives the result at the call site.

\textit{3) Persistent State Tainting:}
Taint persists across transactions via contract storage. Writing tainted values to storage marks slots as tainted; reads from tainted slots propagate taint. The analysis handles simple variables, packed struct members, mappings \texttt{(keccak256(key, p))}, and arrays \texttt{(keccak256(p) + I)}.

\textit{4) Implicit Control-Flow Tainting:}
Taint propagates based on control dependencies. When tainted variables appear in conditionals (textit{if}, textit{require}, textit{assert}), all control-dependent statements are conservatively tainted to maximize detection capability. Note that it may lead to over-tainting as it is based on the principle of minimizing false negatives at the taint analysis phase.

\subsubsection{Taint Sinks}
Taint sinks, corresponding to Phase 3 of our model, represent operations that confer economic benefit. When tainted data reaches a sink, a taint path is reported.
We consider two fundamental pathways for illegitimate profit: direct and indirect value extraction.
Direct value extraction occurs when attackers manipulate price oracles and immediately exploit incorrect calculations to transfer assets directly to their address.
Indirect value extraction involves a two-step process. First, the attacker corrupts the protocol's internal state to credit unearned value. Second, they withdraw the value through a subsequent, seemingly legitimate transaction. Our analysis focuses on the critical first step, state corruption.
\cref{fig:taint_sinks} shows the three types of taint sinks defined by us.
\yepang{Briefly explain why considering these three types is sufficient.}\lu{Updated.}

\begin{figure}[t!]
\centering
\small
\begin{align*}
& \frac{\operatorname{Call}(cs, f, \_, \_) \wedge \operatorname{Transfer}(r,a) \wedge \operatorname{Arg}(cs, \_, a) \wedge a\downarrow}{\operatorname{Sink}(cs)}
\tag{\textbf{Ether and token transfer}} \label{sinks:ether-transfer}
\\[1ex]
& \frac{\neg \operatorname{IsPublic}(f) \wedge \exists \sigma, v.s.t. [\operatorname{SSTORE}(\sigma, v) \in f \wedge \operatorname{IsMappingSlot}(\sigma)]}{\operatorname{IsLedgerUpdate}(f)}
\\[1ex]
& \frac{\operatorname{Call}(cs, f, args, \_) \wedge \operatorname{IsLedgerUpdate}(f) \wedge args\downarrow}{\operatorname{Sink}(cs)}
\tag{\textbf{Internal ledger update}} \label{sinks:inter-ledger-update}
\\[1ex]
& \frac{\operatorname{SSTORE}(\sigma,v) \wedge \operatorname{IsMappingSlot}(\sigma) \wedge v\downarrow}{\operatorname{Sink}(\operatorname{SSTORE}(\sigma, v))}
\tag{\textbf{Economic state write}} \label{sinks:economic-state-write}
\end{align*}
\caption{Definition of taint sinks.}
\label{fig:taint_sinks}
\end{figure}

\textit{1) Tainted Ether and Token Transfer.}
We define sinks for low-level value transfer calls (\texttt{.call}, \texttt{.send}, \texttt{.transfer}) where the amount derives from a tainted source. Additionally, we flag calls to an ERC20 token's \texttt{transfer()} or \texttt{transferFrom()} function as sinks when the amount argument is tainted.
As formalized in Formula~\ref{sinks:ether-transfer}, this captures direct fund drainage.
In \cref{lst:zzf_burn}, the \texttt{burnFeeRewards} function is marked as a taint sink as it is a value-transfer function.


\textit{2) Tainted Internal Ledger Update.}
Beyond native currency, DeFi protocols manage value through internal accounting systems or ledgers. The most common pattern is \texttt{mapping(address $\Longrightarrow$ uint)}, which maintains critical financial data such as balances and shares.
Attackers can manipulate oracles to inflate balances for later withdrawal. We flag calls to ledger-update functions with tainted values (Formula~\ref{sinks:inter-ledger-update}). Functions are classified as ledger-update if they are \texttt{internal} or \texttt{private} and modify mapping state variables.

\textit{3) Tainted Economic State Write.}
While many ledger updates are abstracted and captured by Sink 2, some may be performed via direct, low-level state writes. 
For direct state writes, we flag \texttt{SSTORE} operations writing tainted values to economic balance storage locations (Formula~\ref{sinks:economic-state-write}). This detects two-step exploits where attackers corrupt internal state without immediate fund extraction.

\subsubsection{Taint Analysis Algorithm}
Next, we explore all possible execution paths that could lead to price manipulation. The algorithm consists of three main phases, as detailed in Algorithm~\ref{alg:taint-analysis}.
\input{sections/algo}
The analysis begins with the \AlgPreProcess procedure (Line 2), which parses the input smart contract \AlgContract to construct Control Flow Graphs (\AlgCFGs), which serve as the foundational data structures for applying propagation rules.
Next, the algorithm performs a fixed-point computation to identify all tainted variables.
The \AlgIdentifySources procedure (Line 3) initializes the \AlgTaintMap by identifying all initial taint sources according to our definitions.
Then, the algorithm enters a repeat-until loop (Line 4-7). In each iteration, the \AlgPropagate function is called to apply our data-flow and control-flow propagation rules, expanding the \AlgTaintMap with newly discovered taints. The loop continues until no new taints are introduced in an iteration, signifying that a fixed point has been reached.
Finally, with the complete \AlgTaintMap, the \AlgIsSink procedure (Line 10) inspects every instruction \AlgInstruction to determine if it represents an economic sink. If a sink is found, \AlgReconstructPath (Line 11) traces the taint dependencies from the sink back to its source, and the resulting taint path is added to the output set for further analysis (Line 12).
For example, the ZZF protocol (\cref{lst:zzf_burn}) exhibits a taint path that propagates data directly from the external price oracle \texttt{uniswapRouter.getAmountsOut()} to the value-transfer function \texttt{burnFeeRewards}.




\subsubsection{Path Grouping and Abstraction}
The taint analysis phase may generate numerous paths that are syntactically different but semantically redundant. To address this computational expense and reduce noise, we introduce a path grouping and abstraction phase.
Our approach is based on the insight that paths sharing the same fundamental vulnerability characteristic can be treated as a single analysis unit.
We transform each raw taint path into an enriched data structure by parsing IR instructions to extract high-level semantic features. As shown in Formula~\ref{eq:critical_implication}, we identify security-critical operations: external calls and state variable writes.
\begin{equation}
\label{eq:critical_implication}
\frac{op = \operatorname{EC}(cs, f, \dots) \lor op = \operatorname{SSTORE}(\sigma, v)}{\operatorname{IsCritical}(op)}
\end{equation}
For each path $P$, we generate a unique group key $\operatorname{Key}(P)$ (Formula~\ref{eq:key_definition}) comprising the source function, sink function, and critical operations $op$. Paths sharing the same key are clustered into group $G_k$ (Formula~\ref{eq:group_definition}).
\begin{equation}
\label{eq:key_definition}
\operatorname{Key}(P) \coloneqq \left\langle \operatorname{Source}(P), \operatorname{Sink}(P), \left\{ op \mid op \in P \land \operatorname{IsCritical}(op) \right\} \right\rangle
\end{equation}
\begin{equation}
G_k \coloneqq \left\{ P \mid \operatorname{Key}(P) = k \right\} \label{eq:group_definition}
\end{equation}
Each group is consolidated into a representative case $P_r(G_k)$ (Formula~\ref{eq:pr_definition}) by selecting the longest path. The rationale is that the longest path is more likely to capture the most complex control flow and include the logic present in the shorter, subsumed paths. The structured summary serves as input to the LLM phase.
\begin{equation}
\label{eq:pr_definition}
P_r(G_k) \coloneqq \operatorname*{argmax}_{P \in G_k} |P|
\end{equation}
\yepang{You may formally define this. Critical operations also need a clearer explanation.} \lu{Updated.}

\subsection{LLM-based Reasoning}
To effectively leverage LLMs for identifying price manipulation vulnerabilities from suspicious paths collected during taint analysis, a sophisticated prompting strategy is needed. We observe that a single, static prompt is insufficient, as it fails to guide the LLM's reasoning process in a context-aware manner, frequently leading to inaccurate conclusions.
To address this, we propose a two-stage, template-driven prompting strategy that dynamically customizes LLM input based on the specific details of each taint path. Our methodology operates on the principle of providing each LLM agent with only the information most relevant to its specific objective.
The process begins with the path filtering stage, which filters a large volume of taint paths to isolate a small set of plausible candidates. Filtered paths are then passed to the attack simulation stage, which attempts to construct a concrete, step-by-step exploit scenario using these filtered paths.
For each stage, we design prompt templates containing placeholders. A pre-processing script parses the raw taint analysis output and populates these placeholders, generating customized and detailed prompts for the LLM. This ensures that the model's analysis is tailored and dynamically adapts to the specific characteristics of the DeFi protocol under review.

\begin{figure}
    \includegraphics[width=\linewidth]{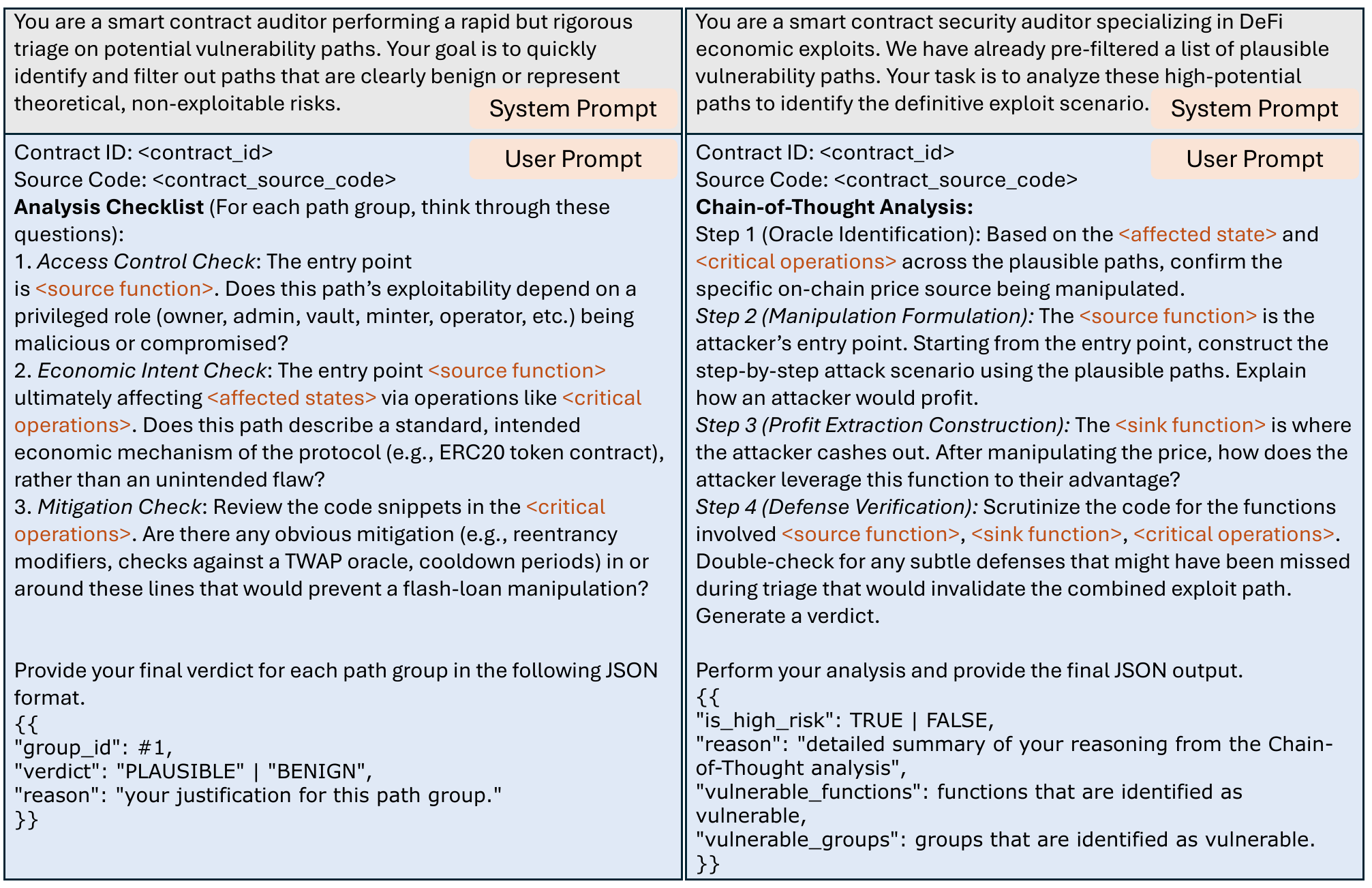}
    \caption{Prompt Templates of the Path Filtering Stage (left) and the Attack Simulation Stage (right).}
\label{fig:prompts}
\end{figure}

\subsubsection{Path Filtering Stage}
The path filtering stage serves as a heuristic filter. It examines taint paths, eliminating those incompatible with the attack model's premises to focus on plausible candidates.
We create a benign behavior checklist derived from the attack model, comprising validation rules selected based on price manipulation vulnerability characteristics and existing research findings~\cite{wang2024defiguard, xie2024defort, wen2024foray}.
Each rule evaluates whether a code path meets necessary conditions for successful attacks.

The left of Figure~\ref{fig:prompts} presents the prompt template for path filtering, which configures the LLM as a smart contract auditor applying rule-based checks.
Each rule contains placeholders for specific taint information, including \texttt{<source function>}, \texttt{<sink function>}, \texttt{<affected states>}, and \texttt{<critical operations>}.
The \texttt{<source function>} represents the entry point where an external attacker initiates the manipulation attack.
The \texttt{<sink function>} is the terminal point where the manipulated state is consumed to generate profit for the attacker.
The \texttt{<critical operations>} includes all low-level external call functions that participate in the execution path between the source and sink functions. These operations represent the core mechanisms through which state manipulation is propagated throughout the system.
The \texttt{<affected states>} refers to state variables whose value is tainted by the source function and subsequently used by the sink function.
The template includes guided analysis questions: (1) \textit{Access Control:} Does the exploit require compromised privileged accounts? (2) \textit{Economic Intent:} Is this an intended economic feature of the protocol, not an unintended flaw? (3) \textit{Mitigation:} Does the code have mitigations like oracle checks or cooldowns to prevent manipulation? 
\yepang{Can we present the three sets of rules using bullets?} \lu{Updated.}
For each path group, we populate this template and concatenate the instances into a complete LLM prompt.

\subsubsection{Attack Simulation Stage}
After path filtering, this stage examines filtered paths to determine if they can create viable economic exploits. We use the template-and-populate approach to guide the LLM in building plausible exploit scenarios.

The right half of Figure~\ref{fig:prompts} outlines the prompt template positioning the LLM as a security auditor analyzing high-potential paths.
The template structures the task as step-by-step flash loan attack analysis, with a chain-of-thought reasoning guiding the LLM through four steps.
First, the LLM identifies the specific on-chain price source being manipulated. Second, it constructs a detailed attack scenario starting from the entry point, using plausible paths and explaining the attacker's profit mechanism. Third, it examines the cash-out method by analyzing how an attacker would leverage the profit function. Fourth, it performs a final defense check, re-evaluating all involved functions for subtle, overlooked mitigation like TWAP or reentrancy guards.
The LLM outputs structured JSON containing verdict, vulnerable functions/paths, and attack explanation.
By populating this template with key functions, operations, and state variables from plausible taint paths, we guide the LLM through the logical exploit sequence. This structured approach enhances the LLM's ability to synthesize taint data into plausible attack narratives.

\subsection{Semantic Sanity Checker}
While the two-step LLM approach effectively reasons about attack feasibility on taint paths, it may still generate false negatives when failing to recognize subtle mitigation controls within the code.
To address this, we introduce a semantic sanity checker as the final filtering stage to validate LLM results.
This checker serves as a high-precision semantic filter verifying whether high-risk paths are neutralized by established defense mechanisms~\cite{xie2024defort, wu2023defiranger}.\yepang{Are there any empirical evidences from literature to support our heuristic rules? Are the three defense mechanisms common? Can they cover most cases?}\lu{Updated}

Based on the structured Intermediate Representation (IR) of the contract, the checker enables control flow and data dependency analysis.
We leverage Slither~\cite{feist2019slither} to parse Solidity code into structured IR format. The checker executes heuristic-based filtering to prune false positives based on defensive code patterns.
The most common cause of LLM false positives is failure to recognize security controls that block exploits. Our checker analyzes the call graph of vulnerable functions flagged by the LLM, searching for defense mechanisms that protect high-risk paths:

1) \textit{Privilege-based defense}: we check if external function calls, especially those interacting with common, well-established DEXs (e.g., Uniswap V2) or manipulating core contract business logic, are protected by access control modifiers.
This is achieved by identifying common modifiers (e.g., \texttt{onlyOwner}, \texttt{onlyAdmin}) and \texttt{require(msg.sender}==\texttt{owner)} patterns that restrict execution to privileged accounts.

2) \textit{Temporal defense}: to counter economic exploits like flash loan attacks, developers often introduce time-based controls. Our checker identifies these patterns by searching for \texttt{require} and \texttt{revert} statements that enforce a delay between actions (e.g., \texttt{require(block.timestamp >= lastActionTime + cooldownPeriod)}).

3) \textit{Fee-on-transfer pattern}: we recognize the fee-on-transfer token mechanism~\cite{feeontransfer} as contextually benign.
This pattern involves swap operations within transfer functions where balance updates are finalized before external calls, adhering to Checks-Effects-Interactions~\cite{checkseffectsinteration}.
Our tool verifies balance updates precede external interactions and calls target known DEX routers, distinguishing legitimate fee collection from malicious reentrancy patterns.

%% file: sections/algo.tex
\begin{algorithm}[t!]
    \DontPrintSemicolon
    \caption{\small Taint Analysis Algorithm}
    \label{alg:taint-analysis}
    \small
    \Input{\AlgContract, the smart contract to be analyze}
    \Output{\AlgTaintPaths, a set of taint paths }
    
    \BlankLine
    
    \Fn{\AlgTaintAnalysis{\AlgContract}}{
        \AlgCFGs $\gets$ \AlgPreProcess(\AlgContract)\; \nllabel{line:preprocess}
        \AlgTaintMap $\gets$ \AlgIdentifySources(\AlgContract)\; \nllabel{line:dentifySources}
        
        \Repeat{\AlgTaintMap == \AlgOldTaintMap}{ 
            \AlgOldTaintMap $\gets$ \AlgTaintMap\;
            \AlgTaintMap $\gets$ \AlgPropagate(\AlgTaintMap, \AlgCFGs)\;
        } 
        
        \AlgTaintPaths $\gets \varnothing$\;
        
        \ForEach{instruction \AlgInstruction in \AlgContract}{
            \If{\AlgIsSink(\AlgInstruction, \AlgTaintMap)}{
                \AlgPath $\gets$ \AlgReconstructPath(\AlgInstruction, \AlgTaintMap)\;
                \AlgTaintPaths $\gets$ \AlgTaintPaths $\cup \{\AlgPath\}$\;
            }
        }
        \Return{\AlgTaintPaths}\;
    }
\end{algorithm}

%% file: sections/5_Evaluation.tex
\section{Evaluation}
We aim to evaluate the following research questions:
\begin{itemize}[topsep=0pt, leftmargin=*]
    \item \textbf{RQ1 (Effectiveness):} How effective is \texttt{PMDetector} at detecting price manipulation vulnerabilities in real-world smart contracts?
    \item \textbf{RQ2 (Comparison with SOTA):} How does \texttt{PMDetector} perform compared with the state-of-the-art tools on price manipulation vulnerability detection?
    \item \textbf{RQ3 (Ablation Study):} How does each component of \texttt{PMDetector} contributes to its overall performance?
    \item \textbf{RQ4 (Execution Cost):} What are the execution time and token usage costs of \texttt{PMDetector}?
\end{itemize}

\subsection{Experiment Setup}
\noindent\textbf{Dataset.}
To evaluate the effectiveness of \texttt{PMDetector}, we constructed a comprehensive ground-truth dataset with two components: a vulnerability set ($D_{vul}$) and a non-vulnerability set ($D_{safe}$).
The $D_{vul}$ set consolidates cases from four academic benchmarks~\cite{xie2024defort, chen2024flashsyn, wu2023defiranger, wu2024strengthening} and the widely adopted industry-maintained DeFiHackLabs incident repository~\cite{defiincidentlist}.
From DeFiHackLabs, we collected all confirmed price manipulation incidents reported between January 2023 and May 2025.
We constructed the $D_{vul}$ set using the following criteria:
\begin{itemize}[topsep=0pt, leftmargin=*]
\item \textbf{Open-source and peer-reviewed}: The selected vulnerability cases must be open-source and widely used in academic papers or real-world industries.
\item \textbf{Clear vulnerability classification}: Each case must be explicitly labeled with price manipulation vulnerability identifiers, ensuring taxonomic clarity and consistency.
\item \textbf{Verifiable exploit impact}: The security implications of each vulnerability must be clearly demonstrated through Proof of Concept (PoC) or documented in formal security audit reports.
\end{itemize}

Based on the above criteria, we examined these five sources to establish our evaluation benchmark for \texttt{PMDetector}. Two of the authors participated in this process and manually verified the datasets.
We identified 73 vulnerable contracts with price manipulation vulnerabilities across these datasets, with Solidity versions ranging from 0.4 to 0.8, and established our analysis benchmark. The average lines of code (LoC) for contracts in the benchmark is 301.3.
For the non-vulnerable set $D_{safe}$, we sampled 288 of the 1,195 protocols from the DeFiTainter dataset~\cite{kong2023defitainter}.
All involved protocols have been audited by security agencies and are reported to be free of price manipulation vulnerabilities. The sample size corresponds to a 95\% confidence level with a 5\% margin of error~\cite{barlett2001}.

\noindent\textbf{Implementation.}
We implemented the \texttt{PMDetector} prototype in Python with approximately 3,000 lines of code. For taint analysis and sanity checker, we utilized the Slither Intermediate Representation (SlithIR) ~\cite{feist2019slither}.
For the LLM evaluation, we selected three state-of-the-art Large Language Models (LLMs) based on their code understanding capabilities and cost-effectiveness: GPT-4.1~\cite{gpt4.1}, Gemini-2.5-Flash~\cite{gemini2.5flash}, and Qwen3-235B-A22B~\cite{qwen235b}.

\noindent\textbf{Evaluation Metrics.}
For the experiments, we employed five metrics, including true positive (TP), false negative (FN), false positive (FP), precision, recall, and F1-score, to evaluate their performance. For the RQ4 (Execution Cost), we evaluated the execution time, the input and output tokens consumed, and the financial costs of LLMs.

\subsection{RQ1: Effectiveness}
We evaluate \texttt{PMDetector}'s effectiveness by measuring its true positive(TP), false negative (FN), false positive (FP), precision, recall, and F1-score. Table~\ref{tab:model_performance} shows the performance.
Out of the 73 vulnerable cases, we successfully converted 68 of them into SlithIR and constructed the taint graph. The result is based on these 68 valid cases.

\begin{table}[t!]
  \centering
  \caption{Performance comparison of \texttt{PMDetector} across different models.}
  \small
  \label{tab:model_performance}
  \begin{tabular}{l|rrr|rr|r}
    \toprule
    \textbf{Model} & \textbf{\#TP} & \textbf{\#FN} & \textbf{Recall} & \textbf{\#FP} & \textbf{Precision} & \textbf{F1-Score} \\
    \midrule
    Gemini2.5-flash     & 60 & 8 & 0.88 & 6 & 0.90 & 0.90 \\
    GPT-4.1              & 57 & 11 & 0.84 & 0 & 1.00 & 0.91\\
    Qwen3-235B-A22B     & 59 & 9 & 0.87 & 10 & 0.86 & 0.86\\
    \bottomrule
  \end{tabular}
\end{table}

\noindent\textbf{Overall Results.}
As detailed in Table~\ref{tab:model_performance}, \texttt{PMDetector} demonstrates high effectiveness across all three LLMs. It achieves a recall of up to 88\% (with Gemini2.5-flash) and a precision of up to 100\% (with GPT-4.1). The results indicate that our approach is highly capable of accurately identifying vulnerable code paths while maintaining a low false positive rate.
The results also highlight a classic trade-off. GPT-4.1 achieves perfect precision by being more conservative, while Gemini2.5-flash and Qwen3-235B-A22B achieve higher recall at the cost of a small number of false positives. Even with models that produce FPs, the precision remains high (90\% for Gemini and 86\% for Qwen). This demonstrates the robustness of our methodology across different language model architectures.

\noindent\textbf{False Negatives.}
False Negatives occur at both the static analysis and LLM-based reasoning stages. In taint analysis, false negatives arise when highly modular architectures with complex inter-contract call chains break taint propagation. For instance, in Mahalend~\cite{mahalend}, taint paths originating in \texttt{Pool.sol} cannot be traced to sinks deep within library contracts accessed via delegatecall.
The LLM-based reasoning stage introduces additional false negatives by incorrectly discarding valid vulnerability paths for two primary reasons. First, overly strict filtering based on access control leads the LLM to dismiss paths originating from privileged functions, such as \texttt{\_setComptroller} or \texttt{\_reduceReserves} in GoodCompound~\cite{goodcompound}. Second, insufficient reasoning about complex economic logic causes the LLM to default to assuming that syntactically correct financial formulas are economically sound, failing to identify subtle vulnerabilities like ImperVexV3~\cite{impermaxv3}.

\noindent\textbf{False Positives.}
The majority of false positives generated by \texttt{PMDetector} stem from the LLM-based reasoning stages, mainly due to two reasons. 
First, the framework misclassifies non-price-manipulation bugs as its target vulnerability class. For instance, in Inverse~\cite{inversefinancefiRM}, \texttt{PMDetector} flags a critical typo (\texttt{add996} instead of \texttt{add96}) that causes transfers to fail. This is a Denial of Service (DoS) vulnerability; however, the LLM-generated explanation attempts to frame it as a price manipulation attack.
Second, the LLM inadequately assesses the role of access control mechanisms.
The Path Filtering stage is designed to evaluate defensive measures, yet it can still fail to recognize that restricting a function to a trusted owner mitigates public attacks. The Attack Simulation stage then proceeds to construct narratives that assume the attacker has gained privileged access, thus leading to false positives.

\begin{answertorq}
\textbf{Answer to RQ1:}
\texttt{PMDetector} is effective at detecting price manipulation vulnerabilities, achieving a recall of up to 88\% (with Gemini2.5-flash) and a precision of up to 100\% (with GPT-4.1). GPT-4.1 achieves the best overall performance with an F1-score of 0.91.
\end{answertorq}

\subsection{RQ2: Comparison with SOTA}
\textbf{Baselines.}
We evaluate our method against two state-of-the-art tools: DeFiTainter~\cite{kong2023defitainter} (static analysis) and GPTScan~\cite{sun2024gptscan} (LLM-based analysis). We also implement \texttt{Pure-CoT}, which provides LLMs with only contract source code, task description, and chain-of-thought prompts without additional domain-specific enhancements.
We exclude on-chain approaches~\cite{wu2023defiranger, wang2024defiguard, xie2024defort, wen2024foray} requiring blockchain transaction data, attack synthesis approaches~\cite{bosi2025following} targeting attack contracts, and closed-source tools~\cite{wang2024smartinv, gao2025airaclex, wu2024strengthening} to maintain evaluation fairness.
DeFiTainter analyzed 60 cases (13 failed due to compilation errors). GPTScan requires single consolidated files; we successfully flattened 68 out of 73 contract projects to meet this requirement. 
To ensure fair comparison across all LLM-based approaches, we use GPT-4.1~\cite{gpt4.1} as the base language model for GPTScan, \texttt{Pure-CoT}, and \texttt{PMDetector}.

\begin{table}[t!]
  \centering
    \caption{Performance comparison with existing tools.
}
  \label{tab:baseline_comparsion}
  \begin{threeparttable}
  \small
  \renewcommand{\arraystretch}{0.8}
  \begin{tabular}{p{10em}|rrr|rr|r}
    \toprule
    \textbf{Tool} & \textbf{\#TP} & \textbf{\#FN} & \textbf{Recall} & \textbf{\#FP} & \textbf{Precision} & \textbf{F1-Score} \\
    \midrule
    DeFiTainter~\cite{kong2023defitainter}  & 4  & 56 & 0.06 & 0 & 1.00 & 0.13 \\
    GPTScan~\cite{sun2024gptscan}\tnote{1}       & 34 & 34 & 0.50 & 44 & 0.44 & 0.47 \\
    \midrule
    Pure-CoT\tnote{1}         & 40 & 28 & 0.59 & 10 & 0.80 & 0.68 \\
    PMDetector\tnote{1}      & \textbf{57} & \textbf{11} & \textbf{0.84} & \textbf{0} & \textbf{1.00} & \textbf{0.91}\\
    \bottomrule
  \end{tabular}
  \begin{tablenotes}
  \small
      \item[1] We use GPT-4.1~\cite{gpt4.1} as the base model for GPTScan, Pure-CoT, and PMDetector.
    \end{tablenotes}
  \end{threeparttable}
\end{table}

\noindent\textbf{Overall Results.}
Table~\ref{tab:baseline_comparsion} presents the comparison results with all baseline tools.
DeFiTainter achieves zero false positives with pre-defined patterns but low recall (0.06) due to rigidity, resulting in an F1-score of 0.13.
GPTScan improves significantly with 0.50 recall and 0.47 F1-score, but low precision (0.44) yields 44 false positives. This suggests that its predefined scenarios are too general and its static confirmation phase is insufficient to filter out the LLM's incorrect reasoning.
\texttt{Pure-CoT} surpasses GPTScan with 0.59 recall and 0.80 precision through step-by-step reasoning, but generates 10 false positives likely due to ``hallucinating'' infeasible attack paths. Besides, it generates 28 false negatives, which suggests that without domain-specific guidance, the LLM can still fail to comprehend highly complex or subtle vulnerability logic.
The results demonstrate that \texttt{PMDetector} significantly outperforms both traditional static tools and LLM-based methods, demonstrating its effectiveness in detecting price manipulation vulnerabilities.

\begin{answertorq}
\textbf{Answer to RQ2:}
\texttt{PMDetector} demonstrates superior performance compared to existing static analysis tools and LLM-based methods.
\end{answertorq}

\subsection{RQ3: Ablation Study}
To investigate the contribution of the core components of \texttt{PMDetector}, we
compare the performance of \texttt{PMDetector} and its variants.
We create four variants of \texttt{PMDetector},
\ie,
(a) \texttt{PMDetector-NoFP}, which omits the LLM-based path filtering stage;
(b) \texttt{PMDetector-NoAS}, which omits the LLM-based attack simulation stage;
(c) \texttt{PMDetector-NoLLM}, which removes all LLM modules, including both path filtering and attack simulation;
(d) \texttt{PMDetector-NoSC}, which skips the final semantic sanity checker that validates high-risk taint paths.
Gemini2.5-flash~\cite{gemini2.5flash} is the base model in this RQ.

\begin{figure*}[t!]
    \centering
    \includegraphics[width=5.2in]{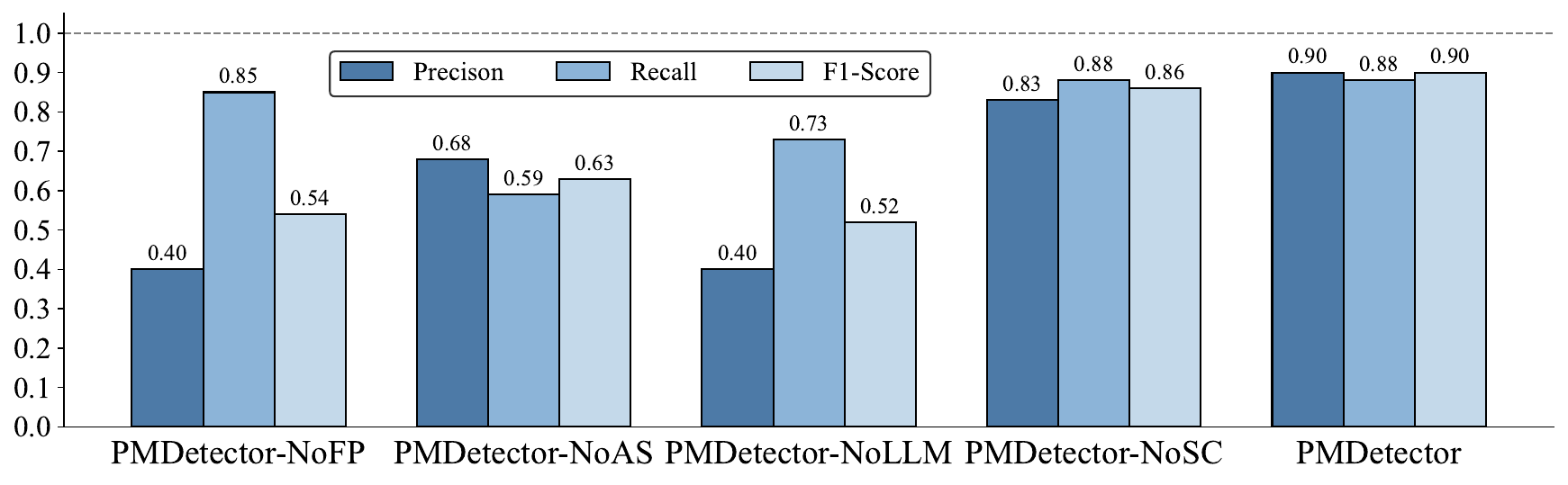}
    \caption{Ablation result of \texttt{PMDetector}.}
    \label{fig:ablation_result}
\end{figure*}

\noindent\textbf{Overall Results.} 
The performance of the variants is shown in Figure~\ref{fig:ablation_result}. Removing the LLM path filtering stage causes precision to drop from 0.90 to 0.40 due to an over-approximate static taint analysis that generates many false positives. Without the LLM's pruning ability, the system is overwhelmed by these paths. Removing the LLM attack simulation stage leads to a recall drop from 0.88 to 0.59, highlighting its importance in confirming true vulnerabilities by simulating attacks on high-risk paths. Removing both LLM modules results in the worst performance, with the F1-score dropping to 0.52. Omitting the sanity checker causes a moderate drop in precision (from 0.90 to 0.83) and F1-score (from 0.90 to 0.86), but recall remains unchanged at 0.88, showing that the sanity checker filters false positives without discarding true positives. These results confirm that both LLM modules and the sanity checker are crucial for effective vulnerability detection.

\begin{answertorq}
\textbf{Answer to RQ3:}
The ablation study confirms that all components of \texttt{PMDetector} are essential for optimal performance.
\end{answertorq}

\subsection{RQ4: Execution Cost}
In this RQ, we evaluate the execution cost of \texttt{PMDetector} across different LLMs.
Following existing practices~\cite{sun2024gptscan}, we use tiktoken~\cite{tiktoken}, a tokenization tool from OpenAI, to estimate the token counts for all models.
We measure several key metrics:
(a) \textit{Time}: total execution time between issuing a request and receiving the result,
(b) \textit{Avg. Time}, average execution time per contract,
(c) \textit{Avg. In} and \textit{Avg.Out}, average number of input and output tokens consumed per contract, respectively,
and
(d) \textit{Avg. Cost}, average financial cost to identify one price manipulation vulnerability .

\begin{table}[!t]
  \centering
  \caption{Running time and financial cost of \texttt{PMDetector}.}
  \label{tab:overhead}
  \renewcommand{\arraystretch}{0.8}
  \small
  \begin{tabular}{l|rrrrr}
    \toprule
    \textbf{Model} & \textbf{Time (s)} & \textbf{Avg. Time (s)} & \textbf{Avg. In}  & \textbf{Avg. Out} & \textbf{Avg. Cost (USD)} \\
    \midrule
    Gemini2.5-flash     & 3067.6 & 9.7 & 11,325.0 & 1342.0 & 0.016 \\
    GPT-4.1              & 1398.4 & 4.0 & 10,834.8 & 1044.3 & 0.030\\
    Qwen3-235B-A22B     & 1734.9 & 4.9 & 10,829.1 & 847.6 &0.002 \\
    \bottomrule
  \end{tabular}
\end{table}

\noindent\textbf{Overall Results.}
As shown in Table~\ref{tab:overhead}, there are significant differences in performance and cost.
In terms of speed, \texttt{PMDetector} powered by GPT-4.1 is the most time-efficient, requiring only 4.0 seconds per contract on average. Qwen3-235B-A22B is also efficient, with an average time of 4.9 seconds. In contrast, Gemini2.5-flash is the slowest, taking 9.7 seconds per contract.
Regarding financial cost, Qwen3-235B-A22B is the most economical, costing a mere \$0.002 per vulnerability. This is substantially lower than both Gemini2.5-flash (\$0.016) and GPT-4.1 (\$0.030).
We observe that token consumption (Avg. In and Avg. Out) is relatively stable across models, indicating that the cost difference stems primarily from the pricing structures of the API services rather than the verbosity of the models' outputs.
Compared with human audits, which typically cost \$5,000–\$15,000~\cite{auditcost}, all LLMs are dramatically more affordable.
These findings highlight a clear trade-off between speed and cost. While GPT-4.1 offers the fastest analysis, Qwen3-235B-A22B provides a practical and scalable solution, achieving comparable speed at a fraction of the financial cost.

\begin{answertorq}
\textbf{Answer to RQ4:}
Compared to traditional human audits, which average roughly \$10,000, \texttt{PMDetector} offers a substantially more efficient and cost-effective alternative (\$0.030 per vulnerability using GPT-4.1). Overall performance varies with the selected LLM back end.
\end{answertorq}



%% file: sections/6_Discussion.tex
\section{Discussion}
\subsection{Case Study}
We demonstrate practical usage of \texttt{PMDetector} by analyzing a real-world vulnerable contract (Listing~\ref{lst:zero_day}, with identifiers anonymized for privacy).
The contract facilitates token swaps using an external AMM as a price oracle. The vulnerability stems from the contract's reliance on the AMM's spot price, which is calculated from live, manipulable token reserves (\texttt{ORACLE.getReserves()} in the \texttt{\_cal\_swap\_out} function). An attacker can exploit this by: (1) securing a flash loan to artificially skew the AMM's reserves, manipulating the price, and (2) immediately calling the \texttt{executeSwap} function, which reads the distorted price and executes a swap at an extremely favorable rate. We have submitted this case to an auditing platform for validation.

\begin{figure}[t!]
\begin{lstlisting}[xleftmargin=2em,]
function _cal_swap_out(uint amountIn) internal view returns (uint amountOut) {
    (uint reserveA, uint reserveB) = ORACLE.getReserves();
    uint numerator = amountIn * reserveB;
    uint denominator = reserveA + amountIn;
    amountOut = numerator / denominator;
}

function executeSwap(uint amountIn) external returns (uint amountOut) {
    amountOut = _cal_swap_out(amountIn);
    TOKEN_A.transferFrom(msg.sender, address(this), amountIn);
    ORACLE.executeSwapOnBehalfOf(msg.sender, amountIn, amountOut);
}
\end{lstlisting}
\caption{A zero-day price manipulation vulnerability.}
\label{lst:zero_day}
\end{figure}

\subsection{Threats To Validity}
\textbf{Reproducibility.} LLM behavior is not fully deterministic. Although we employ structured prompts and chain-of-thought reasoning to guide the model, variations in version, internal state, or prompt changes could affect reproducibility. To support reproducibility, we have released our dataset, source code, and relevant materials in our GitHub repository.

\noindent\textbf{Training Data Contamination.}
The LLM may have encountered vulnerabilities from our evaluation dataset during training, potentially inflating performance metrics. To mitigate this, we tested a baseline \texttt{Pure-COT} that provided a naive CoT prompt without our framework's structural guidance. This baseline achieved only 0.59 recall and 0.68 F1-score, demonstrating that even with potential prior exposure, the LLM cannot reliably identify vulnerabilities. We further validated our approach's effectiveness by using \texttt{PMDetector} to discover previously unknown vulnerabilities, several of which were submitted to auditing platforms for confirmation.

\noindent\textbf{Scope and Extensibility.} Our framework targets flash-loan-based price manipulation attacks, focusing on the most prevalent and financially devastating attack vectors in DeFi.
However, the core workflow of \texttt{PMDetector} is extensible. The key idea is the hybrid methodology, which leverages static taint analysis to identify potentially vulnerable data flows and employs LLMs to reason about their complex economic exploitability, followed by a final validation step. This pipeline can be extended to other smart contract vulnerabilities requiring deep semantic understanding.

\noindent\textbf{Detection Limitations.} 
Our taint analysis relies on common smart contract patterns. Unconventional or obfuscated contracts may evade detection. Our analysis focuses on direct profit extraction through asset transfers and ledger corruption, potentially missing subtle exploitation mechanisms.
The semantic sanity checker and the checklist in the LLM path filtering stage also rely on heuristics for defensive patterns, potentially missing novel defenses (false positives) or approving flawed implementations. We mitigate this through rules based on established security practices like Checks-Effects-Interactions~\cite{checkseffectsinteration}.

\noindent\textbf{Exploit Generation.} \texttt{PMDetector} provides detailed vulnerability reports but does not generate runnable exploit scripts. This limitation stems from the complexity of price manipulation attacks, which depend heavily on the dynamic on-chain state that is difficult to model statically. 
Exploiting such vulnerabilities requires prerequisites like precise knowledge of external liquidity pool reserves and real-time oracle prices.
Future work could integrate \texttt{PMDetector} with mainnet forking environments to provide a realistic on-chain context (token prices, pool balances) needed for symbolic execution or fuzzing of the transaction sequences.

%% file: sections/7_RelatedWork.tex
\section{Related Work}
\subsection{Smart Contract Vulnerability Detection}
The automated detection of vulnerabilities in smart contracts has been extensively studied, with approaches broadly categorized into static, dynamic, and machine learning-based methods.
Static analysis techniques~\cite{feist2019slither, luu2016making, so2020verismart, tikhomirov2018smartcheck, xue2020cross} inspect source code or bytecode without execution, examining code structure and semantics to identify potential vulnerabilities.
Dynamic analysis techniques~\cite{choi2021smartian, jiang2018contractfuzzer, nguyen2020sfuzz, mossberg2019manticore} execute smart contracts to observe runtime behavior, monitoring execution paths and state changes to detect vulnerabilities.
Machine learning-based methods~\cite{wang2020contractward, xu2021novel, gao2019smartembed, cai2023combine, liu2021combining} leverage various feature representations, such as bytecode and data-flow dependencies, to train models that can generalize across different contract implementations.
The recent success of LLMs in code understanding has opened new avenues for smart contract security analysis~\cite{sun2024gptscan, wei2024llm, ma2024combining, boi2024smart}. GPTScan~\cite{sun2024gptscan} combines LLM code analysis capabilities with traditional static analysis to confirm potential vulnerabilities. 
LLM-SmartAudit~\cite{wei2024llm} introduces a conversational framework with specialized agents that collaborate to analyze contracts.
iAudit~\cite{ma2024combining} employs a two-stage fine-tuning process with separate Detector and Reasoner models.

\subsection{Price Manipulation Vulnerability Detection}
Research in detecting price manipulation vulnerabilities can be categorized into three categories: static analysis, dynamic analysis, and LLM-based analysis.
Static analysis tools~\cite{kong2023defitainter, wu2024strengthening, bosi2025following} inspect contract code to find vulnerabilities before exploitation. DeFiTainter~\cite{kong2023defitainter} uses inter-contract taint analysis to track manipulated price inputs. SMARTCAT~\cite{bosi2025following} analyzes newly deployed bytecode in real-time to identify malicious contracts.
Dynamic analysis techniques~\cite{wu2023defiranger, wang2024defiguard, xie2024defort, wen2024foray} analyze transaction data to detect malicious financial activity. DeFiRanger~\cite{wu2023defiranger} constructs cash flow trees from transactions and detects attacks using predefined patterns. FORAY~\cite{wen2024foray} uses a Domain-Specific Language to model financial operations and generate attack sketches.
LLM-based analysis~\cite{sun2024gptscan, wang2024smartinv, gao2025airaclex, zhong2025defiscope} has emerged as a promising direction. GPTScan~\cite{sun2024gptscan} breaks down vulnerability types into code-level scenarios and properties and using an LLM to match them, followed by a static confirmation step. SmartInv~\cite{wang2024smartinv} employs a Tier of Thought (ToT) prompting strategy to infer crucial security invariants.


%% file: sections/8_Conclusion.tex
\section{Conclusion}
In this paper, we introduced \texttt{PMDetector}, a novel hybrid framework that addresses the limitations of existing tools in detecting DeFi price manipulation vulnerabilities. Our approach combines static taint analysis for comprehensive path identification, a two-stage LLM pipeline for semantic reasoning and attack simulation, and a final static checker to validate results and mitigate hallucinations. 
Evaluated on a comprehensive benchmark of 73 real-world vulnerable contracts, \texttt{PMDetector} achieves state-of-the-art performance with 88\% precision and 90\% recall, significantly outperforming previous static analysis and LLM-based methods. At just \$0.03 and 4.0 seconds per audit, our work demonstrates that the structured integration of static analysis techniques and large-scale language models provides a powerful, scalable, and economically viable paradigm for securing smart contracts against complex economic attacks.